# Changes in charge density vs changes in formal oxidation states: The case of Sn halide perovskites and their ordered vacancy analogues

Gustavo M. Dalpian,[1,2,*] Qihang Liu,[1,*] Constantinos C. Stoumpos,[3,*] Alexios P. Douvalis,[4] Mahalingam Balasubramanian,[5] Mercouri G. Kanatzidis,[3,6] Alex Zunger[1]

[1]Renewable and Sustainable Energy Institute, University of Colorado, Boulder, Colorado 80309, USA
[2]Centro de Ciências Naturais e Humanas, Universidade Federal do ABC, Santo André, SP, Brazil
[3]Materials Science Division, Argonne National Laboratory, Argonne, Illinois 60439, USA
[4]Physics Department, University of Ioannina, 45110 Ioannina, Greece
[5]Advanced Photon Source, Argonne National Laboratory, Argonne, IL 60439, United States
[6]Department of Chemistry, Northwestern University, Evanston, Illinois 60208, USA
[*]These authors contributed to the work equally.

**Abstract**

Shifting the Fermi energy in solids by doping, defect formation or gating generally results in changes in the charge density distribution, which reflect the ability of the bonding pattern in solids to adjust to such external perturbations. In the traditional textbook chemistry, such changes are often described by the formal oxidation states (FOS) whereby a single atom type is presumed to absorb the full burden of the perturbation (change in charge) of the whole compound. In the present paper, we analyze the changes in the position-dependence charge density due to shifts of the Fermi energy on a general physical basis, comparing with the view of the FOS picture. We use the halide perovskites $CsSnX_3$ (X=F, Cl, Br, I) as examples for studying the general principle. When the solar absorber $CsSnI_3$ (termed 113) loses 50 % of its Sn atoms, thereby forming the ordered vacancy compound $Cs_2SnI_6$ (termed 216), the Sn is said in the FOS picture to change from $Sn^{2+}$ to $Sn^{4+}$. To understand the electronic properties of these two groups we studied the 113/216 compound pairs $CsSnCl_3/Cs_2SnCl_6$, $CsSnBr_3/Cs_2SnBr_6$, and $CsSnI_3/Cs_2SnI_6$, complementing them by $CsSnF_3/Cs_2SnF_6$ in the hypothetical cubic structure for completing the chemical trends. These materials were also synthesized by chemical routes and characterized by X-ray diffraction, $^{119}$Sn-Mössbauer spectroscopy and K-edge X-ray Absorption Spectroscopy. We find that indeed in going from 113 to 216 (equivalent to the introduction of two holes per unit) there is a decrease in the *s* charge on Sn, in agreement with the FOS picture. However, at the same time, we observe an *increase* of the *p* charge via downshift of the otherwise unoccupied *p* level, an effect that tends to replenish much of the lost *s* charge. At the end, the change in the charge on the Sn site as a result of adding two holes to the unit cell is rather small. This effect is theoretically explained as a 'self-regulating response' [H. Raebiger, S. Lany, and A. Zunger, Nature **453**, 763 (2008)] whereby the system re-hybridizes to minimize the effect of the charge perturbation created by vacancy formation. Rather than having a single pre-selected atom (here, Sn) absorb the full brunt of the perturbation producing two holes/cell, we find that the holes are distributed in a complex pattern throughout the octahedral systems of $X_6$ ligands, forming hole orbitals with some specific symmetries. This clarifies the relation between FOS and charge transfer that can be applied to a wide variety of materials.

**I. Introduction**

The halide perovskites[1] $A^I M^{IVB} X^{VII}_3$ (with A = organic or inorganic monovalent cation; $M^{IVB}$ = metalloid Sn, Pb; and $X^{VII}$ = Cl, Br, I) are successful as semiconducting *solar absorbers* [band gaps as low as 1.2 eV for $CH_3NH_3SnI_3$ and 1.5 eV for $HC(NH_2)_2PbI_3$]. This is surprising, given that these compounds represent in essence *ionic halides* $A^+[(Pb/Sn)X_3]^-$. Indeed, analogous ionic compounds $A^+[Ba/SrI_3]^-$ have far larger band gaps (~ 4 eV for $CsSrI_3$), being insulators and useless as photovoltaic (PV) absorbers. We will see that we owe this special feature of PV-useful *small gap, formally ionic* $A^I M^{IVB} X^{VII}_3$ compounds that have recently climbed to the top of the thin-film PV charts to the existence of bonding-viable latent *p* orbitals in the Sn but not in alkaline earth elements such as Sr/Ba.

The solar absorbers $A^+[(Sn)X_3]^-$ (termed 113) compounds can be stabilized against loss of Sn, an oxidizing process that results ultimately in the 'ordered vacancy compound' $A_2Sn\square X_6$ (termed "216" compound, with 50% Sn vacancies[2,3]) where A is a positive cation (Cs or organic molecule). The "proximity effect" of the unoccupied p orbitals to the bonding arena in $CsSnX_3$ -- which allows such compounds to have low, PV-useful band gaps -- raises another interesting question regarding the role of the Sn valence electrons as one transforms $CsSnX_3$ to $Cs_2Sn\square X_6$: will Sn (IV) in the 216 compound (formally with configuration $s^0p^0$) also have effective s-p valence electrons as in Sn (II) (formally $s^2p^0$ in the 113 compound, but unlike Sr/Ba(II) in $CaSnO_3$(formally Sn(IV) $s^0$)? This question is interesting because as is generally known, formal oxidation states (FOS) such as Sn(II) and Sn(IV) do not imply complete removal of II or IV valence electrons[4], respectively, yet whatever residual physical valence charge density resides on Sn can control its electronic properties, including its solar band gaps. The broader question here is the general relationship between FOS and the physical valence charge density residing on bonded atoms when the compound as a whole loses charge.

To address these questions, we synthesized and characterized several sets of $CsSnX_3$ and $Cs_2SnX_6$ compounds (X = Cl, Br, I), as well as the reference binary compounds $SnI_2$ and $SnI_4$ (see Supplemental Material [5] for details of solution synthesis methods). These materials were characterized by X-ray diffraction at the Advanced Photon Source at Argonne National Laboratory. We also performed [119]Sn-Mössbauer spectroscopy and K-edge X-ray Absorption Spectroscopy (XAS) to gauge the trends in s-electron charge on the nucleus. By using the density functional theory (DFT), we have also calculated the electronic structure of these 113/216 compound pairs, complementing them by $CsSnF_3$/$Cs_2SnF_6$ in the hypothetical cubic structure for completing the chemical trends in that structure. The calculations were performed by using the Density Functional Theory (DFT) (See Supplemental Material [5] for details of DFT calculations). From this extensive set of data, we find that:

(i) The existence of a multi-channel coupling (e.g., occupied s and empty p in the Sn compounds) enables a self-regulation response (SRR) whereby the system can minimize the effect of the external perturbation (removing 50% Sn from $CsSnX_3$) by creating opposite and nearly compensating charge rearrangements between the different channels: reduction of s charge and enhancement of p charge on Sn. As a result, whereas the *formal* oxidation state of Sn is said to change from II in $CsSnI_3$ to IV in $Cs_2SnI_6$, the *physical charge densities* residing on the Sn atom computed by density functional theory (DFT) does not show significant changes.

(ii) The change in s electrons is captured by the Mössbauser isomer shift, in good *quantitative* agreement with DFT calculations of this quantity.

(iii) Calculated differences in 113 and 216 charge densities at constant geometries identify locations in the unit cell where the two holes produced by removal of 50 % of Sn are actually distributed. Rather than having a single pre-selected atom (here, Sn) absorb the full brunt of the perturbation producing two holes/cell, as classically envisioned by the FOS concept, we find that the holes are distributed in a complex pattern throughout the octahedral systems of $X_6$ ligands, forming hole orbitals with some specific symmetry.

(iv) The above noted SRR is active when the cation sublattice is strongly coupled to the anion sublattice, enabling compensating exchange of *s* and *p* carriers. The classic ionic limit where such communication is disabled exists to some extent in the hypothetical cubic $ASnF_3$/$A_2SnF_6$ fluoride where loss of s electrons is not compensated by gain of p electrons. A more nearly ideal weak-coupling (ionic) limit is revealed in $CsSrI_3$ lacking active valence p electrons. DFT calculations show that the charge resides solely in the s channel, resulting in far wider band gaps than in the covalent systems capable of SRR, and thus enabling solar absorption in the latter but not the former.

Our work provides microscopic, quantum-mechanical content of the relation between FOS and actual charge transfer. The concept described here not only provides an alternative thinking of the fundamental physical chemistry process of response to charging in strongly-coupled orbital systems, but also sheds light on a wide variety of systems with significant ion-ligand hybridization, in addition to halide perovskites, such as materials with multivalence and with strong correlation effect.

**II. The broader issue of FOS, charge density and Self-Regulating Response**

One generally needs to discuss *changes* in charges that result from a redox perturbation, not absolute charges. Indeed, it was established in condensed matter physics that while static atomic charges in a solid are ill defined, the *charges displaced by any perturbation* ('charge in transit') are unambiguous.[67]

The concept of FOS of an ion is connected with the idea of charge transfer[8]. It is commonly defined as

$$\text{FOS(ion)} = Q(\text{compound}) - \Sigma Q(\text{ligands}) \tag{1}$$

Here, the first term on the right-hand side is the observable *physical* charge of the system as a whole. For molecules, it equals the net ionic charge, e.g., +1 for $[NH_4]^+$ or -1 for $[BH_4]^-$; for solids such charges are commonly induced by shifting the Fermi level via carrier doping. The second term corresponds to the sum of the presumed phenomenological effective charges on ligands. Its magnitude could be guessed from the relative electronegativities of the ligand vs. central ion. Eq. (1) incorporates, as the first term a physical (observable) net charge, but the second term is a non-observable set of charges whose magnitude, as will be illustrated, lies to a large extent in the eyes of the beholder. Indeed, the determination of $Q(\text{ligand})$ might require sometimes a deeper understanding of the ligand-to-ion bonding than supplied by the respective electronegativities (For example, it is +4 for the hydrogen $H_4$ ligand in $[NH_4]^+$ and −4 for the hydrogen $H_4$ ligand in $[BH_4]$).

Despite this, the FOS axiom is a numerical estimate of the electronic state of matter that has helped in the development of chemistry using simple, intuitive concepts. Using FOS, a chemist can interpret the preferred geometry of a molecule, or why a chemical bond is shorter than another for a set of compounds with different oxidation numbers. In the specific case of Sn halide compounds examined here (CsSn(II)X3/Cs2Sn(IV)X6), as the formal oxidation state increases, the Sn-X bond length decreases. This can be said to be due to an increase in the electronegativity of Sn as its FOS increases.

The discussion above can be posed as a quantum mechanical computational problem:

According to Eq. (1), a change of the electron count [i.e., a change of $Q$(compound)] necessitates that the FOS (ion) of one ion must change. Will the *charge density* around the ion change too? It was theoretically reported early on that, for $3d$ impurities in silicon and GaAs,[9] and for the de-lithiation process of $LiCoO_2$,[10] the physical charge density stayed nearly constant as $Q$(compound) was varied. This is the essence SRR: in a system where metal ions and ligand form coupled quantum states, when the system is perturbed by reducing the compounds charge (shifting the Fermi energy), the ligands re-hybridize self-consistently (including the possibility of atomic displacements), thereby replenishing much of the charge lost by the cation; the physical charge on the cation is therefore largely unchanged, even though the FOS would change by integer numbers. Next we examine the way the physical charge density changes in the $CsSnX_3$ and $Cs_2SnX_6$ compounds when two holes are introduced into the unit cell by creation of extensive Sn vacancies, comparing the result to the FOS rendering of the same event. The availability of X-ray diffraction, as well as [119]Sn-Mössbauer spectroscopy and K-edge XAS provides crucial evidence, supporting the results of the theory.

**III. Considering s electrons on Sn: experimental and theoretical probes**

**A. The Mössbauer spectroscopy and its DFT interpretation**

If the charge on Sn(II) /Sn(IV) were purely due to absence or presence of s-electrons (as in the ionic picture), what would the chemical trends in perturbed charge densities be? To address this question, [119]Sn Mössbauer spectroscopy is used to probe directly the electron population residing in the *s*-obitals (See Supplemental Material [5] for Mossbauer experiments details). This is expressed by the difference in the isomer shift (IS) between a reference material (here $SnO_2$) and the perovskite sample. This is exemplified by the Mössbauer spectra of $CsSn^{II}Br_3$ and $Cs_2Sn^{IV}Br_6$, with both compounds having a similar coordination, adopting a cubic symmetry at room temperature (see Figure 1a). The experimental spectra of $Cs_2SnBr_6$, with a FOS of IV, resonates at IS = 0.83 mm/s while $CsSnBr_3$, with a FOS of II, has a resonance at IS = 3.90 mm/s. These values clearly indicate that the relative s-orbital charge distributions of $SnO_2$ is much closer to $Cs_2SnBr_6$ than to $CsSnBr_3$, thus fully supporting the FOS picture. Similar trends occur in other compounds of the $CsSnX_3$ and $Cs_2SnX_6$ families and the experimental results are listed in Table S1. The results are in excellent agreement with previously reported values.[11,12,13]

Note that we chose not to include experimental data for CsSnCl$_3$[14] and CsSnI$_3$[15] in this table since these compounds have non-perovskite polymorphs (different coordination environments) which may introduce discrepancies in the experimental trends. The Mössbauer spectra of these compounds have been reported previously by Donaldson et al.[16]

Theoretically, the isomer shift arises from the *difference* of electronic wavefunction density at the nuclear site ($\Delta\rho_{Sn}(0)$) of both the target absorber and source. Giving a specific source reference (in this work we use SnO$_2$), the isomer shift of Sn ($\delta$) is expressed as[17]

$$\delta = \frac{2\pi Zce^2}{3E_\gamma} \Delta\rho_{Sn}(0) \cdot \Delta\langle r^2 \rangle, \qquad (2)$$

where $Z = 50$ is the atomic number of Sn, and $c$, $e$, $E_\gamma$ and $\Delta\langle r^2 \rangle$ denote the speed of light, electron charge, the $\gamma$-ray energy and the mean square nuclear radius for the transition, respectively. The coefficient $\Delta\rho_{Sn}(0) = \sum_{occ}(|\psi_A(0)|^2 - |\psi_S(0)|^2)$ counts the electron density (the sum of all occupied wavefunction square) at the nuclear site with $r = 0$.

Figure 1 shows the correlation between the experimentally measured isomer shift (Fig. 1a) and the DFT-calculated electron density difference at the nucleus $\Delta\rho_{Sn}(0)$ on series of Sn based halide compounds. We find a quantitative linear correlation between the two quantities, indicating that DFT is a good predictor of IS. Similarly, good correlation was reported for other Sn-based compounds when referenced to SnO$_2$.[17]

Although $\Delta\rho_{Sn}(0)$ is the accurate quantity to correlate with the IS, a more intuitive (but less rigorous) quantity is the charge *around* a bonded atom, as it has an intrinsic scale (such as < 2 for *s*-electrons). The total and angular-momentum (*l*) decomposed atomic charges are defined and calculated by projecting the periodic wave functions $\psi_n(k)$ (expressed with plane-wave expansion) on spherical harmonics of Sn site *i* and summing over all occupied bands:

$$Q_{i,l} = \sum_{n,k}^{occ} \langle \psi_n(\mathbf{k})|(|l, i = Sn\rangle\langle l, i = Sn|)|\psi_n(\mathbf{k})\rangle, \qquad (3)$$

where $|l, i\rangle$ is the angular momentum eigenstate of *i*-th atomic site. Here we consider only Sn 5s atomic charge ($l = 0$).

Figure 1c shows that the measured $Q_{Sn5s}$. The isomer shift correlates well with the s-like atomic charge around the bonded Sn atom. As expected, the 113 compounds have more s-charge [$Q(Sn) \sim 2$] than the 216 compounds [$0.5 < Q(Sn) < 1$]. Interestingly, one can delineate two separated groups in Fig. 1c, indicated by ellipses: the compounds that are normally associated with a formal oxidation state of 4+ (SnX$_4$, Cs$_2$SnX$_6$) have less s charge than those associated normally with a 2+ FOS ($SnX_2$, $CsSnX_3$ and $CaSnO_3$). If there were no p electrons as suggested by the ionic picture [Sn(II): $s^2p^0$ and Sn(IV): $s^0p^0$], and no charge density outside atoms ('extra-atomic charge') *this result will imply an association of 113 compounds with FOS=II and 216 compounds with FOS=IV.*

**B. X-ray Absorption Near Edge Structure as a direct probe of the effective charge on Sn relative to its metallic state ($Sn^0$).**

The generally accepted methods of determining the oxidation state of metals include Photoemission Spectroscopy (PES), X-ray Absorption Near Edge Structure (XANES). PES, and in particular X-ray Photoemission Spectroscopy (XPS), is a widely used method which probes the M electron shell (3*d* orbitals) in 1-1000eV spectral range and it is invaluable for assessing the FOS of transition metals. In the case of Sn, however, this method has been shown to be quite insensitive, as in this case, as illustrated here, FOS does not take up the ideal value of the s orbital occupation (FOS=II being $s^2$ and FOS=IV being $s^0$); instead, here FOS reflects a range of electron population of the 5s states.[18,19] In other words, the ideal integer FOS do not by themselves reflect the electron distribution in different Sn(II) and Sn(IV) compounds, and an additional, independent factor exists—the actual s electron count. Ultraviolet Photoemission Spectroscopy (UPS), which probes the electrons on top of the valence band (0-20eV) and in principle can provide answers with respect to the FOS of Sn, is very sensitive to the surface of the sample and is prone to vary when measured on different thin-film substrates.[20] In order to ensure that the observed trends arise from the bulk materials alone, all samples were characterized by powder X-ray diffraction (PXRD) to eliminate the possibility of secondary phases (Figures S2-S6).

The obtained XANES spectra (Figure 2) are characterized by the Sn absorption edge corresponds to the ionization of one 1s electron, which occurs at $E_i$ = 29200.4 eV for the Sn metal,[21] and an above edge peak which corresponds to the transition probability of the 1s electron to the empty 5p orbitals ($\Delta l = \pm 1$ selection rule) to form a core-hole/photo-electron pair (exciton). As the FOS of Sn increases so does the ionization energy and as a result the K-edge absorption shifts to higher energies. It is clearly seen from Figure 2 that compounds with a II FOS ($SnI_2$, $CsSnBr_3$) have a similar absorption energy, whereas the IV FOS compounds ($SnI_4$,

Cs$_2$SnX$_6$) have a significantly higher absorption edge energy. However, we can not tell if the increase of the ionization energy reflects a large ( ~ 2$e$) or small reduction of s orbital count .

Interestingly, the absorption edge energy in the IV FOS compounds increases proportionally with the electronegativity of the halide ($\chi_{Cl}$ > $\chi_{Br}$ > $\chi_{I}$) whereas in the II FOS compounds (Figure 2) the absorption edge is insensitive to the electronegativity of the halide. This trend parallels the DFT results in Fig.1c and implies that when the 5$s^2$ lone pair is present the ligands do not affect the $s$ charge whereas its absence in the Sn$^{4+}$ ions allows the electronegativity of the ligand be the charge-determining factor. This observation is also in good agreement with both Mössbauer spectroscopy and DFT calculations, providing an independent confirmation of the electron populations on Sn orbitals.

The IS values of the Sn$^{4+}$ ions with I$^-$, Br$^-$ and Cl$^-$ environments in the Cs$_2$SnX$_6$ series also follow a trend (similar as that found for the "model" SnX$_4$ compounds,[22] where the decrease in IS values is related to the decrease in 5$s$ electron density at the $^{119}$Sn nucleus, due to the decreasing covalency of the Sn-X bond as the p-orbital configuration changes from 5$p$ (I) to 4$p$ (Br) to 3$p$ (Cl).[23] On the other hand, the spectra of II FOS compounds (SnI$_2$, CsSnBr$_3$), which consistently show an IS = 3.90 mm/s irrespective of the halide or the coordination environment. As shown in Figs. 1b and 1c, the FOS II group the different compounds are clustered, while for the FOS IV group they are more sparsely distributed. Compared with FOS II group that has almost all s electrons occupied, FOS IV group has part of the s electrons ionized.

A second trend that appears from the XANES spectra comes from the intensity of the exciton peak as a function of the halide ligand. The intensity of the exciton peak relates directly to the probability of the 5$p$ orbitals to be occupied by the excited photo-electron and it serves as an indirect measure of the occupation of the Sn orbitals in the Sn-X bonding. This quantity directly relates with the electronegativity difference between Sn and X; the higher the difference the more ionic the bond and the more orbitals are accessible to the photo-electron during the 1$s$-5$p$ transition. This trend is confirmed in Fig. 2, where all the compounds displaying a comparable peak intensity for the same halide ligands.

**IV. The changes in p charge on Sn in strong-coupling Iodide systems counteracts the changes in s charge**

To understand the extent to which the ionic limit is obeyed by these Sn-based compounds, we show in Figs. 3a and 3b the DFT-calculated density of states (DOS) of $CsSnI_3$ and $Cs_2SnI_6$, respectively, and the electronic charge density (Fig. 2c) describing some of the DOS peaks that are important to our analysis. This associates the "s charge" $Q_{Sn5s}$ discussed in the previous section with certain energy windows in the DOS, thus providing better insights into the origin of this charge. Not surprisingly, the electronic charge density enclosed in the DOS energy window corresponding to Peak I (in $CsSnI_3$) and II (in $Cs_2SnI_6$) clearly shows **5s charge** in both compounds (Fig. 3c). The 216 compound has less 5*s* charge around the Sn atom, as evidenced also from the Mössbauer data (Fig 1). However, analysis of the Sn **5p charge** for both compounds reveals important differences: comparing the charge density plots (Fig. 3c) of peak I ($CsSnI_3$) to II ($Cs_2SnI_6$) we see *a decrease of 5s charge density around* Sn, but comparing the charge density plots of peaks III ($CsSnI_3$) to IV ($Cs_2SnI_6$) we see *an increase of 5p charge density around the Sn atom*. *Thus, Sn tends to preserve its total charge in the face of a perturbation that reduces its s-charge by removing 50% of the Sn atoms.*

Moving from charge *density* maps (Fig. 3c) to angular dependent atomic charge Eq. (3), and considering the difference in total charge on Sn between $CsSnI_3$ and $Cs_2SnI_6$, we find the difference is surprisingly small. Our calculations show that while the Sn atom in $CsSnI_3$ loses 0.6 s electrons when it is transformed to $Cs_2SnI_6$, it gains back 0.2 p electrons in the occupied state, so the change in total physical charge on the Sn atom is just 0.4 e. While the s charge is reduced upon removal of Sn atoms from the (113) compound (oxidation), non-s charge acts to offset the decrease in s charge so the physical charge on Sn is similar in 113 and 216. Specifically, while in the ionic limit the valence states of Sn are written as Sn(II): $s^2p^0$ and Sn(IV): $s^0p^0$, in the covalent case the valence states are written as Sn(II): $s^2p^m$ and Sn(IV): $s^{2-\Delta 1}p^{m+\Delta 2}$. Accordingly, the charge difference on Sn is $\Delta 1-\Delta 2$, rather than +2 suggested by FOS in the ionic limit. This is a manifestation of the self-regulation response: keeping the physical charge nearly constant and delocalizing the holes in the X ligands. Thus, one does not expect to find a good correlation between the isomer shift and the **total** physical charge on Sn,( See Supplemental Material, table S1, [5] for details on isomer shifts) indicating that the total charge on Sn cannot reflect FOS in covalent compounds. *Such a SSR effect is very general and is seen in other systems as well.*

Previous study focusing on the oxidation state of Sn in $Cs_2SnI_6$ concluded that the oxidation state of Sn in this material should the (II), the same as in $CsSnI_3$.[4] This would occur because the Sn 5s orbitals would be occupied in both compounds. Our calculations show that this is not the case: The Sn 5s orbitals in 216 are half occupied, while in 113 Sn 5s is completely occupied. Also, previous work disregarded the effect of Sn 5p orbitals. The Sn 5p state in 216 has SSR effect, compensating the loss of 5s charge compared with 113 compound. The facts presented here and not considered before are important to get a complete understanding of charge self-regulation in these compounds.

## V. Energy level model that explains the *s vs p* negative feedback in the strong coupling limit of the CsSnI$_3$ and Cs$_2$SnI$_6$ compounds

Going back to the specific case of 113 vs. 216 Sn compounds, we develop an energy level diagram distilled from our DFT calculations that explains the mechanism of the negative feedback (Fig. 3d). The hybridization between the Sn(5*s*, 5*p*) at the center and the octahedron of I5p ions is governed by symmetry considerations. Specifically, while the Sn atom has spherical s states and dumbbell p states, the I$_6$ cage has O$_h$ point group and their p orbitals are split to several representations. Among them, only the states with A$_{1g}$ symmetry (singlet, basis function $x^2+y^2+z^2$) can hybridize with Sn 5*s* states, while I5*p* T$_{1u}$ states (triplets, basis function x, y, z) couple with Sn 5*p* states. The remaining representations of I$_6$ (e$_g$, T$_{1g}$, T$_{2g}$, T$_{2u}$, etc.) simply have no counterpart in Sn to bond to, and are non-bonding. In Fig. 3(a) and (b), the projected DOS of the hybridization between Sn5*s* and I5*p* A$_{1g}$ orbitals, and between Sn5*p* and I5*p* T$_{1u}$ orbitals are shown by red and green highlights, respectively. For each pair of hybridized states, the lower energy part indicates bonding state, and the higher energy part indicates the antibonding state.

Using the calculated DOS we are able to explain the SRR effect by simplifying the projected DOS into a 4-level model (corresponding to the 4 peaks in Fig. 3a and b). As shown in the schematic plot in Fig. 3d, from 113 to 216 the Sn-I coupling becomes stronger because of the 8% shrink in the size of SnI$_6$ octahedra (See Supplemental Material [5] for structural information). For the Sn5*s* and I5*p*-A$_{1g}$ pair, in 113 both the bonding and antibonding states are occupied, so the Sn 5s orbitals are fully occupied. In contrast, the stronger hybridization pushes the antibonding state above the Fermi level, leading to a loss of occupied Sn 5*s* electron. At the same time, hybridization also depopulates some I5p states, which distribute equally in the I$_6$ octahedra, forming A$_{1g}$ hole orbitals. On the other hand, the Sn 5p orbital, while often neglected, accepted electron density from the Iodine ligands and plays a crucial role on compensating the total charge of Sn. The bonding state of Sn 5p and I5p-$T_{1u}$ is occupied for both 113 and 216 compounds, indicating that there is always electron density in the Sn 5p orbitals. Compared with 113, stronger hybridization in the 216 systems mixes even more Sn *5p* orbitals with the occupied bonding state which is predominantly I5*p* based. Consequently, when the antibonding Sn *5s* electrons get depopulated in response to the formation of Sn vacancies, the Sn *5p* character of occupied orbitals is increased, compensating the loss of *s* electrons.

**VI. Charge density deformation maps for the strong coupling iodine compounds: where do the two holes go?**

We compute the change in unit cell electron charge density due to shift in the Fermi level $E_F$ (brought about by oxidation of CsSnX$_3$ through vacancy formation) without using perturbation theory but rather by calculating self-consistently and independently the charge before and after shifting $E_F$. We nevertheless refer to the ensuing change as *perturbation,* resulting from the creation of vacancies without implying that perturbation theory is involved. Considering the results presented above, another question would be: where do the two holes produced by the 113 to 216 transformation reside if the physical charge on Sn does not change much?

Looking at the charge deformation map we can see the redistribution of charges. This is defined as $\Delta\rho = \rho^{113} - \rho^{216} - \rho^{Sn}$, where $\rho^{113}$ and $\rho^{216}$ refers to the total valence charge density of CsSnI$_3$ and Cs$_2$SnI$_6$, respectively. $\rho^{Sn}$ is the charge density of the sublattice of Sn that is removed, and is used to balance the total charge difference. By necessity we use a *fixed geometry* for all the components either in the lattice parameter of the 113, in Fig. 4a, or the lattice parameter of the 216 compound, as show in Fig. S6. Both cases give similar results. The position of the Sn atoms, Sn vacancies and I atoms are indicated and a density profile along a representative line (Fig. 4b) gives a better idea of the amplitude of the differences. Blue areas indicate negative values.

When we go from the 113 to the 216 compounds there is a small but noticeable charge migration from the Sn atom to the Sn-I bond. The small change in the charge around the Sn atom corroborates that the charge sitting on the Sn atom is very similar for both compounds. Owing to the reconfiguration of charges away from the center of atoms, it is also clear that we move from a more ionic (113) to a more covalent (216) compound. *We see how the SRR works – charge is* *redistributed into positive (red) and negative (blue) oscillating domains so that the perturbation in physical charge upon vacancy formation is minimal.*

**VII. Moving along the series from strong-coupling covalent to the weak-coupling ionic limit (I→Br→Cl→F)**

The discussion above on the SRR pertains to systems where there is strong coupling between the ligand subsystem and the metal subsystem as in the $CsSnI_3/Cs_2SnI_6$ cases. We consider next halide compounds with increasing electronegativity by using X = I, Br, Cl and F in $CsSnX_3/Cs_2SnX_6$. The atomic charges $Q_{i,l}$ of Eq. (3) for Sn 5s and 5p orbitals are shown in Figs. 5a and 5b respectively. We find that in the 113→216 perturbation, the Sn atom loses 0.60, 0.85, 1.04 and 1.31 *5s* electrons for X = I, Br, Cl and F, respectively, indicating the increasing ionic component of the Sn-X bonds. In contrast, for X = I and Br, Sn 5p atomic charges provide a compensation to the loss of the *s* electrons through an increase in the quantity of p electrons going from the 113 to the 216 compound, owing to the SSR effect we have described above. The SSR effect is diminishing as one increases the electronegativity of the X element. *For X = F, both Sn s and p components are losing charge, indicating that the two holes are mostly localized on the Sn atom, consistent with the FOS picture.* Clearly, in the extreme ionic limit of this series (X=F), the SRR is almost gone. Indeed, F represents an electronic phase transition relative to X=I, Br, Cl: By looking at the energetic order of the atomic levels related to Sn and the halogen atoms, we find that *while for I, Br and Cl the halogen p orbital is above the Sn5s orbital (as in Fig. 3d), for the fluoride the 2p orbital is below Sn5s*. This qualitative change is illustrated in Fig. S7, where we present the corresponding diagram. This reversed order of energy levels weakens the coupling between Sn5*p* and F2*p* levels, minimizing the SRR effect.

The maintenance of some Sn 5*s* charge even for the F compounds shows that even our most ionic compound cannot be considered fully ionic. It is instructive to look further for a perfect "no SRR" or 'fully ionic' case in this family. The best candidate is obtained by replacing Sn by a divalent cation with totally inactive (high energy) p states such as Ba or Sr. We have simulated $CsSrI_3$ and observed the contribution of the Sr atom to the electronic structure (See Supplemental Material [5] for projected density of states). *We observe that Sr 5s has a very small contribution to the occupied states, and the component of Sr 5p is negligible*. These are indications that the FOS of Sr in CsSrI$_3$ agrees well with the classic FOS picture Sr(II): 5s$^0$5p$^0$. Indeed, it is quite rare to find such almost ideal ionic FOS systems, as some metal-to-ligand coupling is ubiquities.

**VIII. Band gaps of 113 vs 216 halide compounds**

The absorption spectra of the halides show a consistent increase in the band gap as we move from I to Br and to Cl. The increase of the band gap is much larger for the 216 compounds than for the 113 compounds, as shown in the absorption spectra in Fig. 6a. While in CsSnX$_3$ the band gap lie between $E_g$ = 1.3 - 2.7 eV (X = I, Br, Cl), the band gaps of Cs$_2$SnX$_6$ span across the much wider range $E_g$ = 1.3 - 4.4 eV (X = I, Br, Cl), with CsSnI$_3$ and Cs$_2$SnI$_6$, remarkably, starting from the same initial value. Fig. 6b shows the absorption spectra for SnI$_4$ and SnI$_2$.

These trends can be better visualized in a schematic diagram: Figure 7a shows the measured bandgap trends for these compounds, whereas Fig. 7b shows the calculated bandgaps using a hybrid exchange-correlation functional (HSE). The calculated band gaps are smaller than experiment, as expected for this kind of calculations, although the chemical trends illustrated by the vertical arrows showing the differences in gaps, are similar.

The band structures (See Supplemental Material [5] for the band structures) report a direct band gap for all compounds. The band gap of the 113 compound is at R, whereas for the 216 compounds are at Γ. The dispersion in the 216 bands are very small, since SnX$_6$ are isolated in these compounds.

The increase in the band gap for this 113/216 series of halides can be partially understood by considering the transition between 113 to 216 in distinct computational steps that isolate specific effects. There are three effects that are important to understand this effect: (i) the 216 compound is formed by removing Sn atoms from the 113 compound. The formation of these Sn *vacancies* is found to decreases the band gap, by the formation of *impurity-like* bands in the original gap. These *intermediate bands* can be clearly observed in the band structures in Fig. S7. (ii) The *intermediate bands* will change the character of the CBM in the 216 compounds. For 113 the VBM is composed of halide *p* levels, and CBM is mostly composed of Sn *p* levels. For the 216 compounds, the VBM is also composed of halide *p* levels, but the CBM comes from Sn *s* levels. (iii) The lattice parameter of the 216 compounds is smaller than the corresponding 113. This hydrostatic effect will increase the band gap. At the end, the competition between the creation of impurity levels (decrease the band gap) and hydrostatic effects (increase of the bandgap) explain the similar band gap for iodides and much larger bandgap for bromides and chlorides.

## IX. Discussions

There are two general approaches to the discussion of oxidation states: (i) modeling formal oxidation states that are integer numbers, or (ii) focusing instead on different measures of the position-dependent charge density and arguing their relevance to the response of the system. The integer number approach (i) to FOS can be computed from Eq. (1) by modeling $\Sigma Q$(ligands) from a chemical understanding of the underlying bonding. This gives Sn(II) for $ASnX_3$ and Sn(IV) for $A_2SnX_6$, as well as Bi(III) and Bi(V) in $Ba_2[Bi(II) Bi(V)]O_6$. Alternatively, the integer number approach can be computed from the wavefunctions as recently shown by Jiang et al.,[24] who developed an approach to calculate integer oxidation states based on the theory of electric polarization and quantized charge pumping. From their definition, the structural local environment is encoded into the change of polarization, rather than the charge density. For example, both versions of the integer number approach return for $Ba_2Bi_2O_6$ the same two integer FOS Bi(III) and Bi(V). The concept of *integer FOS(ion)* of Eq (1) is sometimes problematic as a useful tool. Indeed, whether the integer FOS represents, or predicts a physical charge residing on a specific atom has a long-term debate starting from Pauling in 1948,[25] and then supported by R. Hoffmann in 2001,[26] claiming that the charge of +6 on iron in $[FeO_4]^{2-}$ and −2 in $[Fe(CO)_4]^{2-}$, as suggested by FOS, is impossible. This was also shown by Sawatzky and coworkers[27] for the perovskite rare-earth nickelates ($RNiO_3$), where Ni behaves like 2+ (rather than 3+) in a normal high-spin state (S = 1), and there is one hole per 3 oxygens in the O $2p$ band of states.

Here we focus instead on approach (ii) where one considers the position dependence of the charge density in response to Fermi level shift. We find that considering any reasonable atomic volume around the central ion gives that the change in charge is very different than what is conveyed by integer number of FOS. This poses questions whether these theories can be predictive or descriptive: the theory of the self-regulating response involves computing the change in self-consistent charge densities of a given system under different Fermi energies (induced by doping, or defect formation, or gating) and observing how metal-ligand re-hybridization, structural distortions and atomic and relaxations conspire specifically to minimize the change in charge density on the central metal ion. Maximal SRR is found in strongly coupled ion-ligand systems ($Cs_2SnI_6$), whereas minimal SRR is found in weakly coupled systems ($Cs_2SnF_6$, see Fig. 5) where the ligand and metal ion orbitals are largely decoupled. Examples of strong SRR include the $LiCoO_2$ system where de-lithation causes a very small change in the oxidation of Co[10] as well as 3d impurities in covalent semiconductors such as Si.[28,29]

The leading prediction of strong calculated SRR pertains to behavior under localization of electrons in solids reflected by the phenomenology of effectively screened Coulomb interaction. Strong SRR implies that the effective Coulomb repulsion $U_{eff}$ is reduced below what conventional linear screening will imply, including the extreme case of "relaxation induced negative U"[30] or "exchange induced negative U"[31]. See also recent discussion in Ref 32. In addition, the fact that the calculated charge density on Sn is almost unchanged predicts that the system can tolerate with impunity large changes in electron count without having the central ion (Sn or transition metal elements) shift their orbital energies by large effective Coulomb energies. Consequently, all physics that should be sensitive to $U_{eff}$ become more inert than in systems with weak SRR. This includes the co-existence of a few charge states of transition metal impurities in the relatively narrow energy window of the band gaps of Si or GaAs[28] ($E_g$ between 1 and 2eV), even though the nominal screened 3d Coulomb energies would suggest that change of the charge state will entail an upwards shift of the levels by $\sim U \gg E_g$. Such predictions are applicable to a broad range of metal-ligand feedback system.

**X. Summary**

This work discussed correlations between formal oxidation states of atoms and the actual charge density residing in that atom. We show, using Sn-based halide perovskites as prototype materials, that a self-regulating response is observed when these materials are subjected to strong perturbations. In our case the perturbation was the removal of 50% of the Sn atoms when we go from $CsSnI_3$ to $Cs_2SnI_6$. The SSR acts in such a way that the charge density residing on the Sn atom in both cases is almost equal, although the formal oxidation state in both changes from Sn(II) to Sn(IV).

Our conclusions are supported by extensive ab initio calculations and experimental measurements including X-ray diffraction, Mossbauer spectroscopy and XANES, showing the nature of the compensating mechanism in these systems: while the Sn s charge decreases when going from $CsSnI_3$ to $Cs_2SnI_6$, the Sn p charge actually increases. The idea of a self-regulating response should span to other systems, making our conclusions useful also for other types of materials.

**Acknowledgements**

The work at the University of Colorado Boulder was supported by the U.S. Department of Energy, Office of Science, Basic Energy Sciences, Materials Sciences and Engineering Division under Grant No. DE-SC0010467 to C.U. Work on photovoltaic relevant absorption characteristics was supported by the U.S. Department of Energy, Energy Efficiency and Renewable Energy, Under the SunShot "Small Innovative Programs in Solar (SIPS)" Project Number DE-EE0007366. Work at Northwestern University was supported by grant SC0012541 from the US Department of Energy, Office of Science. Use of the Advanced Photon Source at Argonne National Laboratory was supported by the Basic Energy Sciences program of the US DOE Office of Science under contract DE-SC0010467. GMD thanks financial support from Brazilian agencies FAPESP and CNPq. This work used resources of the National Energy Research Scientific Computing Center, which is supported by the Office of Science of the U.S. Department of Energy under Contract No. DE-AC02-05CH11231.

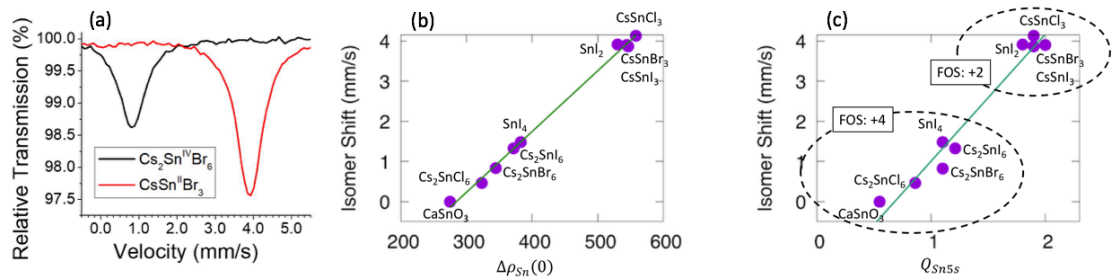

**Fig. 1:** (a) Room temperature $^{119}$Sn Mössbauer spectra spectra of CsSnBr$_3$ and Cs$_2$SnBr$_6$ relative to SnO$_2$. The relative transmission of the spectra is consistent with the concentration of Sn centers in Cs$_2$SnBr$_6$ relative to CsSnBr$_3$ (b) Correlation of experimental $^{119}$Sn isomer shifts (IS) of Mössbauer effect and the valence charge densities at Sn nuclear calculated by DFT. (c) The correlation between measured IS and the DFT calculated *occupied s charge around Sn*. The reference source is CaSnO$_3$, and the green line is the linear fit. The IS of CsSnCl$_3$ and CsSnI$_3$ are taken from Ref. 16, while the others are from our measurements.

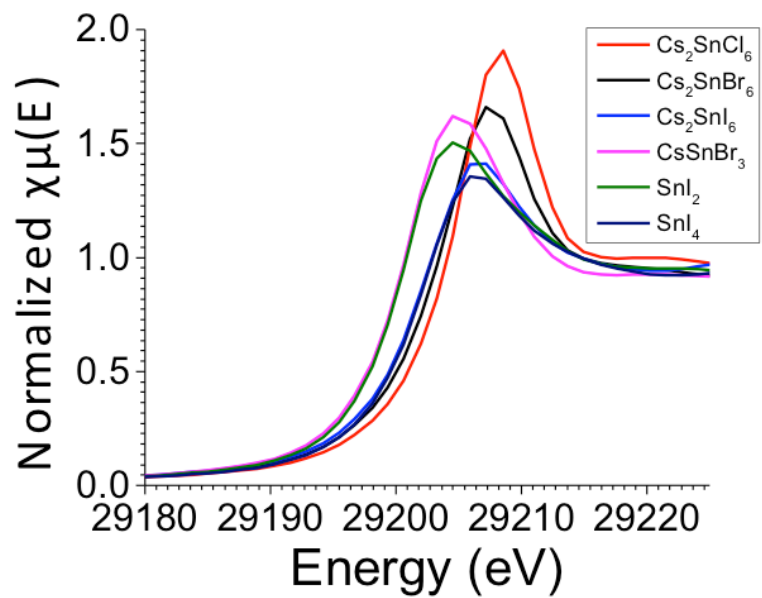

**Fig. 2:** Room temperature Sn K-edge XANES spectra of $CsSnBr_3$, $Cs_2SnX_6$ (X = Cl, Br, I), $SnI_2$ and $SnI_4$.

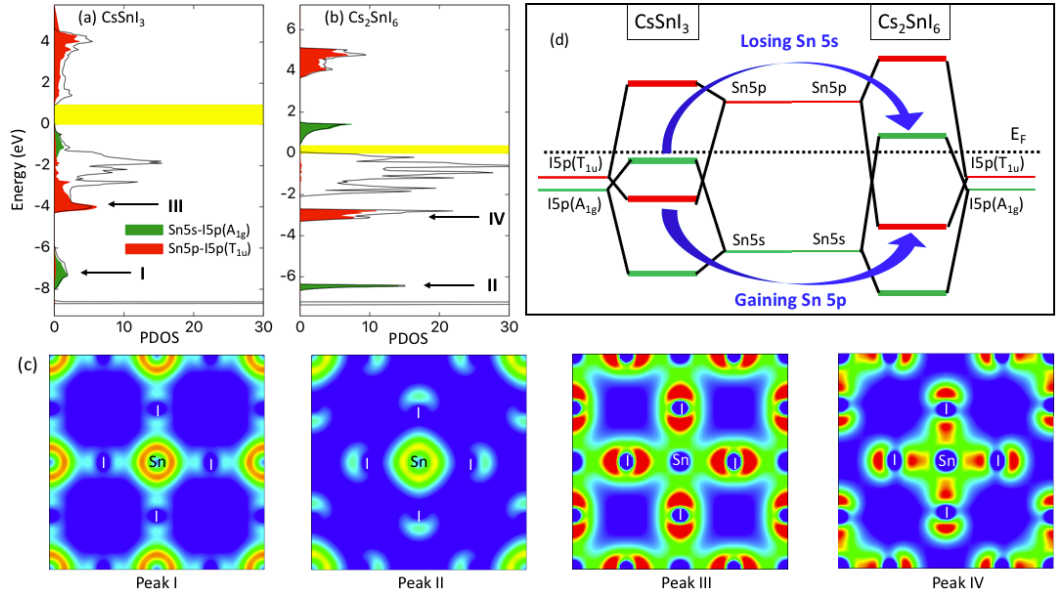

**Fig. 3:** Total density of states (DOS, black line) and its projection on the hybridization states between Sn5$s$ and I5$p$ $A_{1g}$ orbitals, and between Sn5$p$ and I5$p$ $T_{1u}$ orbitals for (a) CsSnI$_3$ and (b) Cs$_2$SnI$_6$. The yellow shadows indicate the band gap. (c) Partial charge density of the DOS peaks shown in (a) and (b). The [100] plane that contains both Sn and I atoms is chosen. Red areas indicate larger charge densities, whereas blue areas indicate smaller charge density. (d) Schematic diagram of Sn5$s$-I5$p$ and Sn5$p$-I5$p$ hybridization and the SSR mechanism for CsSnI$_3$ and Cs$_2$SnI$_6$.

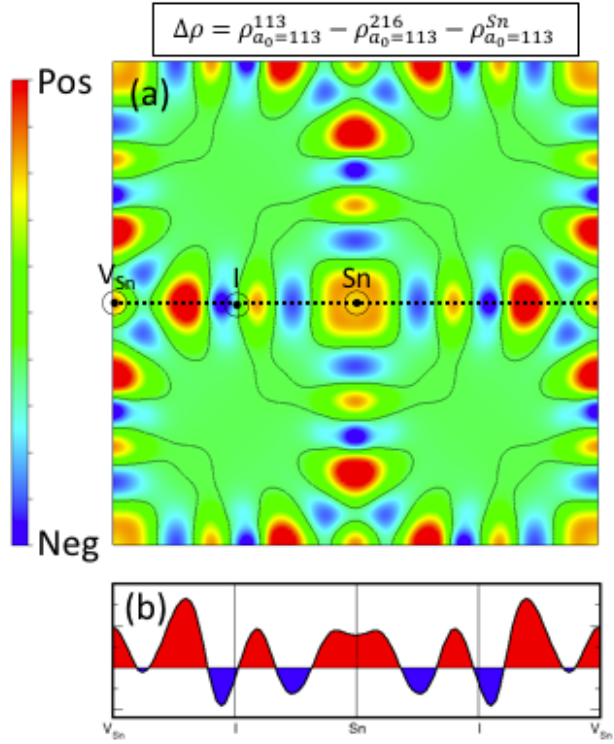

**Fig. 4**: (a) Charge density difference map calculated as $\Delta\rho = \rho^{113} - \rho^{216} - \rho^{Sn}$. The lattice parameter of the 113 compound was used as reference. The charge difference is displayed on the [001] plane containing Sn atoms, Sn vacancies and I atoms. (b) Profile of the charge density difference along the solid line marked in (a) shows the oscillation between charge accumulation and depletion as the function of distance away from the Sn atom. Red and blue areas denote positive and negative charge density difference, respectively.

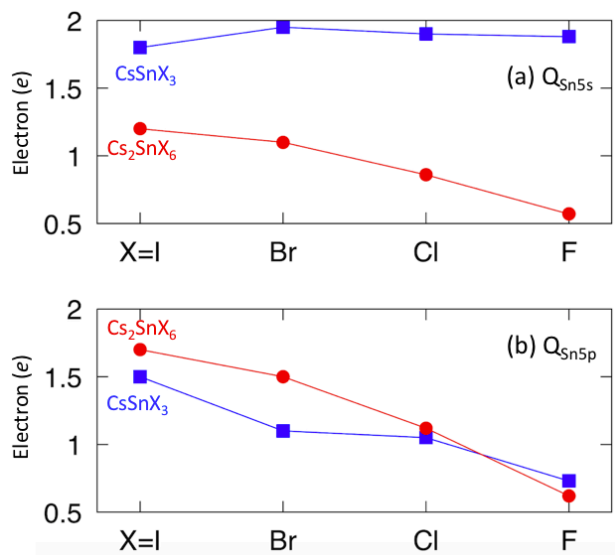

**Fig. 5:** The comparison of (a) 5s and (b) 5p component of Sn valence charge for $CsSnX_3$ and $Cs_2SnX_6$ (X = I, Br, Cl and F).

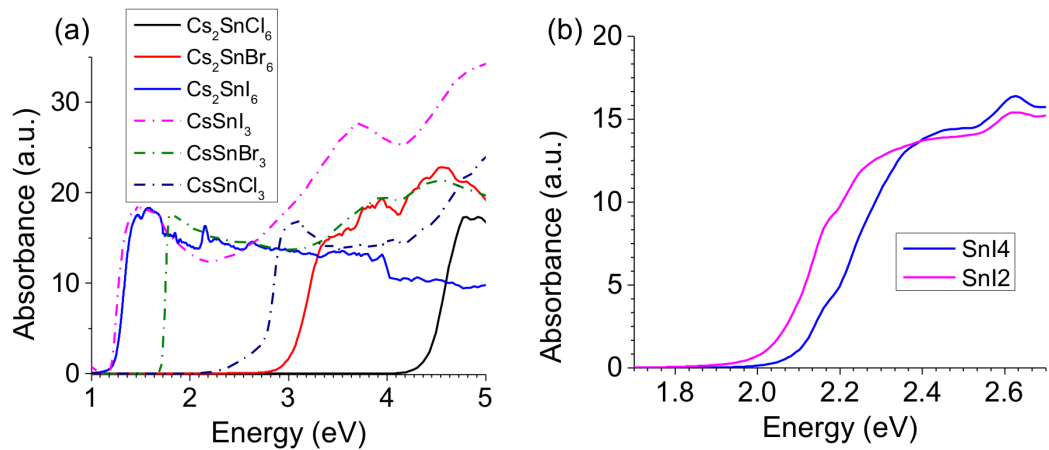

**Fig. 6.** Optical absorption spectra of (a) CsSnBr$_3$ and Cs$_2$SnX$_6$ (X = Cl, Br, I) and (b) SnI$_2$ and SnI$_4$.

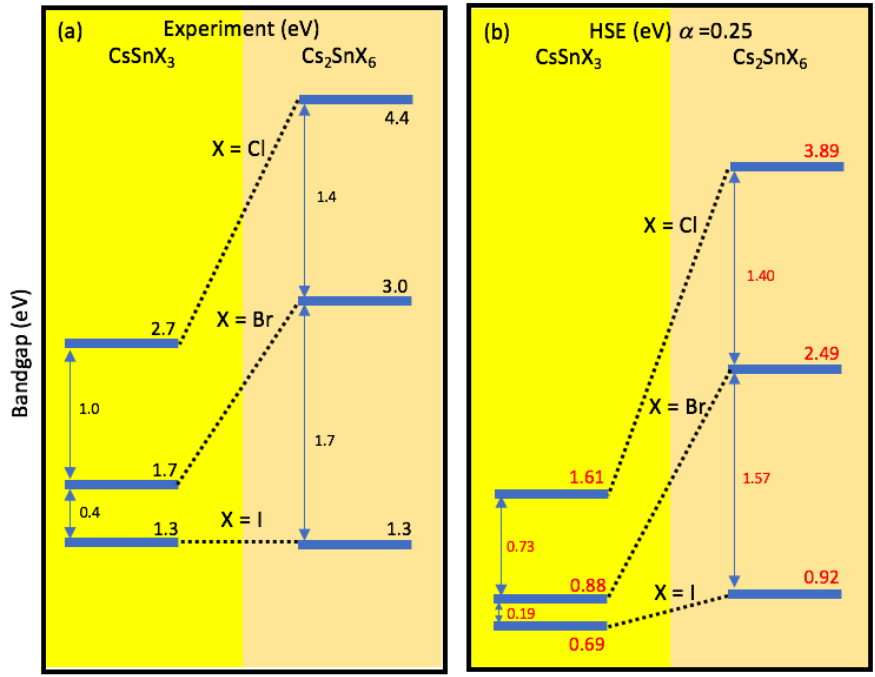

**Fig. 7:** (a) measured bandgaps for 113 and 216 halides. (b) calculated bandgaps using density functional theory and hybrid exchange-correlation functional ($\alpha = 0.25$). Experimental lattice parameters were used in the calculations.

# Supplementary Materials

## Changes in charge density vs changes in formal oxidation states: The case of Sn halide perovskites and their ordered vacancy analogues

Gustavo M. Dalpian,[1,2,*] Qihang Liu,[1,*] Constantinos C. Stoumpos,[3,*] Alexios P. Douvalis,[4] Mahalingam Balasubramanian,[5] Mercouri G. Kanatzidis,[3,6] Alex Zunger[1]

[1]Renewable and Sustainable Energy Institute, University of Colorado, Boulder, Colorado 80309, USA
[2]Centro de Ciências Naturais e Humanas, Universidade Federal do ABC, Santo André, SP, Brazil
[3]Materials Science Division, Argonne National Laboratory, Argonne, Illinois 60439, USA
[4]Physics Department, University of Ioannina, 45110 Ioannina, Greece
[5]Advanced Photon Source, Argonne National Laboratory, Argonne, IL 60439, United States
[6]Department of Chemistry, Northwestern University, Evanston, Illinois 60208, USA
[*]These authors contributed to the work equally.

**S1 Synthesis**

*a) Preparation of the $Cs_2SnX_6$ compounds.* $Cs_2SnX_6$ compounds were prepared following a modification of a reported method.[1] In a typical procedure, $SnO_2$ (3.01g, 20mmol) was dissolved in an excess (20 mL) of the corresponding concentrated, aqueous hydrohalic acid (HCl: 35% w/w; HBr: 48% w/w; HI: 57% w/w) under heating to 110$^o$C. A CsX solution was prepared separately by dissolving $Cs_2CO_3$ (6.52g, 20mmol) in 10 mL of the corresponding acid. Mixing the two solutions under vigorous stiring resulted in the immediate precipitation of a fine powder ($Cs_2SnCl_6$: white; $Cs_2SnBr_6$: tan; $Cs_2SnI_6$: black). The reaction was continued for a further 10 min following the addition of deionized $H_2O$ (5 mL) to facilitate the stirring of the dense mixture and ensure completion of the reaction. The solid was subsequently filtered under vacuum and washed copiously with absolute EtOH followed by drying at 60$^o$C under vacuum. Yield >95% (manipulative losses).

*b) Preparation of the $CsSnBr_3$.* $CsSnBr_3$ was prepared following a modification of a reported method.[2] In a typical procedure, stoichiometric ratios of $SnBr_2$ (20 mmol, 5.570 g) and CsBr (20 mmol, 4.256 g) were placed in a fused $SiO_2$ tube, evacuated at $10^{-3}$ mbar and flame-sealed. The silica tube was placed in a furnace and heated to 600 °C over over 6 h, held at this temperature for 3 h, followed by cooling down to ambient temperature over 3 h, yielding quantitatively a shiny black ingot.

*c) Preparation of the $SnI_2$ and $SnI_4$ compounds.* $SnI_2$ was prepared following a modification of a reported method.[3] In a typical procedure, an excess of granulated Sn (50g) and $I_2$ (70g) were combined in 400 mL of a 2M aqueous HCl under a $N_2$ blanket. Upon completion of the reaction, the hot yellow solution was slowly brought to ambient temperature where $SnI_2$ crystallizes as long red needles after standing for 1 day. Yield ~60-70% (based on $I_2$). Crude $SnI_2$ can be isolated quantitatively by evaporation of the supernatant liquid to dryness. $SnI_4$ was prepared by comination of Sn (5.94g, 50 mmol) and $I_2$ (25.38g, 100mmol) in boiling toluene (100 mL) under reflux. The reaction mixture is heated to boiling under reflux conditions. After completion of the reaction the solution is cooled to ambient temperature to precipitate orange octahedral crystals of $SnI_4$. The crystals are filtered and dried thoroughly under vacuum. Yield ~70%. Crude $SnI_4$ can be isolated quantitatively by evaporation of the supernatant liquid to dryness.

## S2 Experimental Characterization
### a) Determination of the crystal structure parameters.

Powder X-ray diffraction (PXRD) data for $Cs_2SnX_6$ were collected in the $2\theta$ = 5-140° on a Rigaku Miniflex 600 equipped with a Cu Kα source (λ = 1.5406 Å) operating at 40kV and 15 mA. High-resolution temperature dependent powder X-ray diffraction data for $CsSnBr_3$ and $CsSnI_3$ were collected at Advance Photon Source (11-BM). The data were refined using JANA2006[4]. Fig S1 shows the crystal structures of all samples and figs S2-S5 show the PXRD patterns.

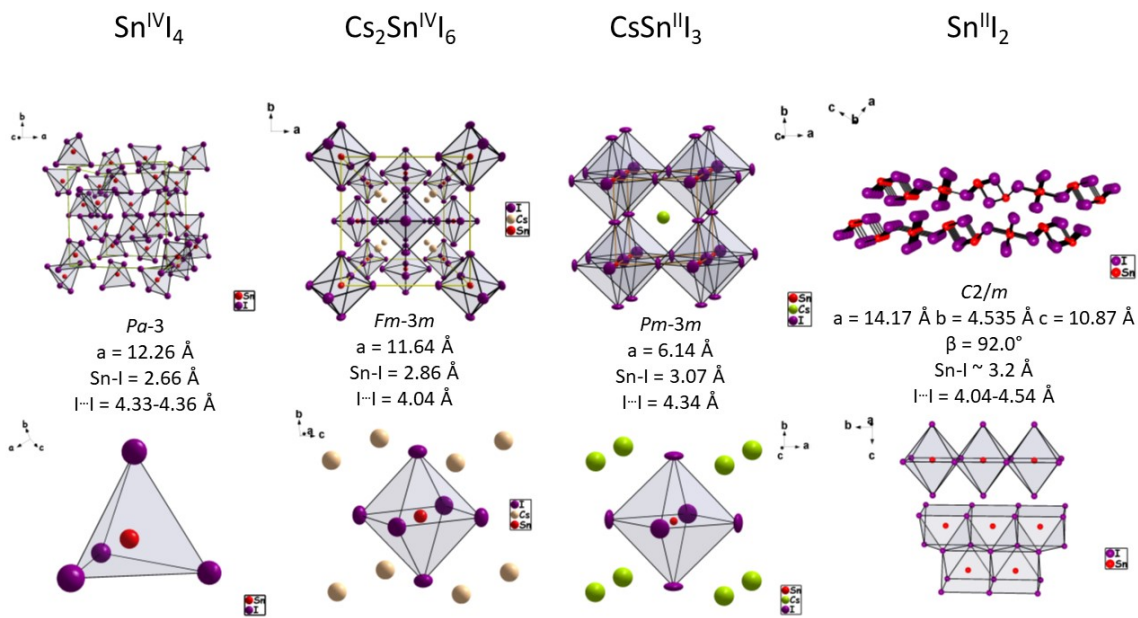

**Figure S1.** Crystal structures of (top) SnI$_4$, Cs$_2$SnI$_6$, CsSnI$_3$ and SnI$_2$ and (bottom) the corresponding building blocks of the inorganic lattices.

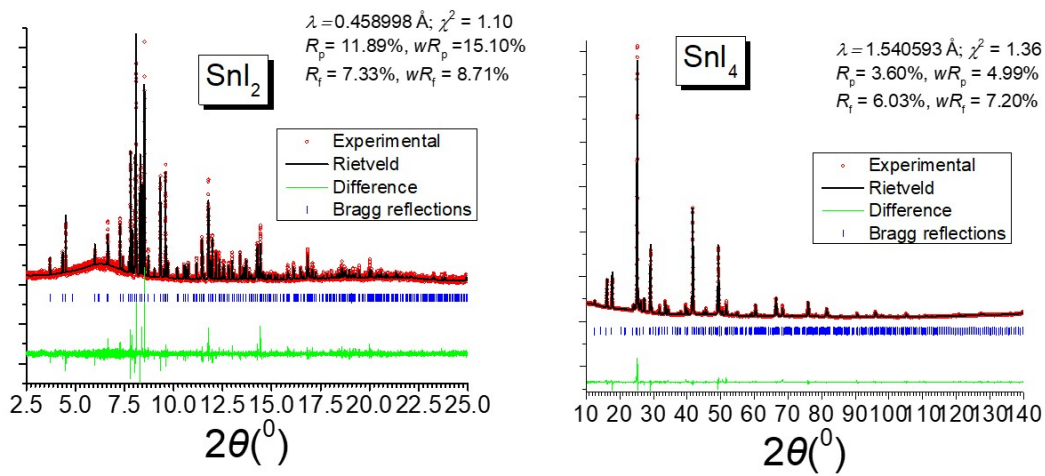

**Figure S2.** PXRD patterns of SnI$_2$ and SnI$_4$ at room temperature.

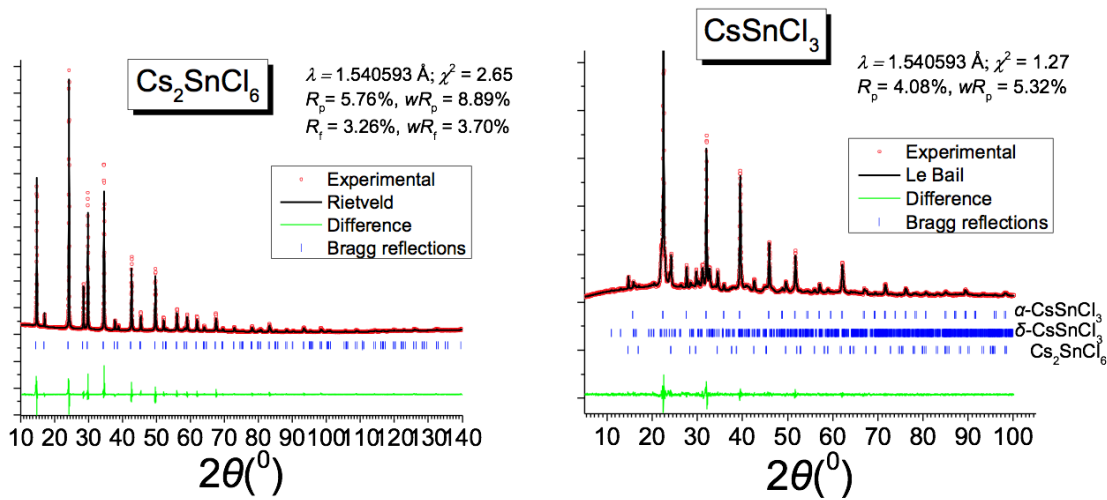

**Figure S3.** PXRD patterns of $Cs_2SnCl_6$ and $CsSnCl_3$ at room temperature. The latter is metastable and transitions from the metastable $α$-$CsSnCl_3$ phase (cubic perovskite) to the stable $δ$-$CsSnCl_3$ phase (monoclinic phase). Partial oxidation to $Cs_2SnCl_6$ also occurs gradually.

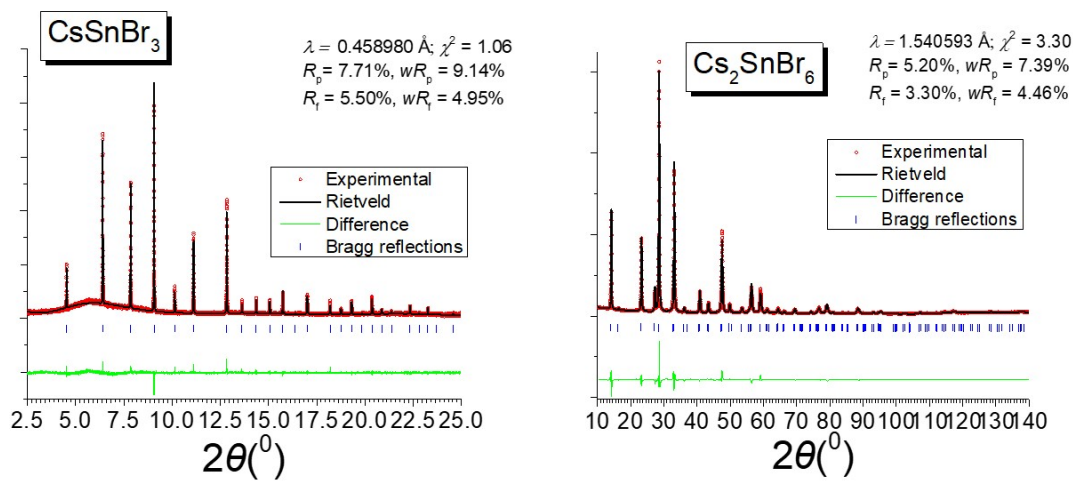

**Figure S4.** PXRD patterns of $CsSnI_3$ (500K) and $Cs_2SnI_6$ (293K).

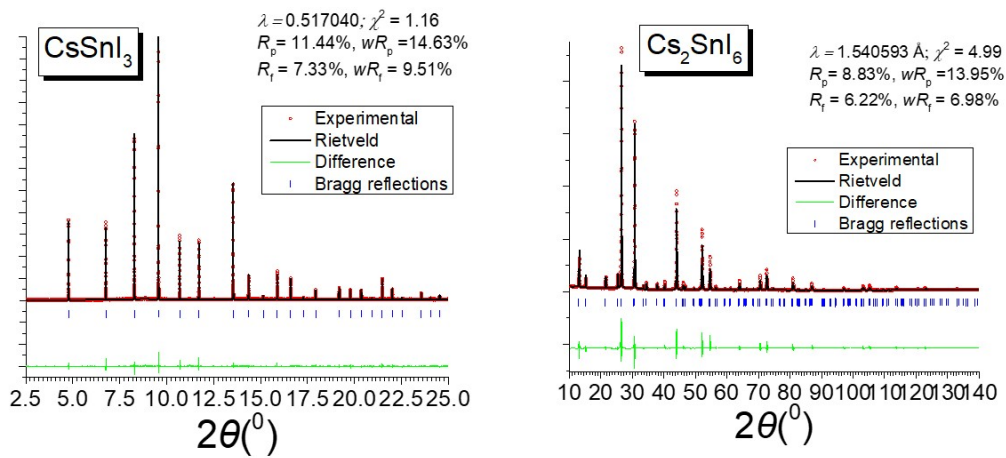

**Figure S5.** PXRD patterns of CsSnBr$_3$ and Cs$_2$SnBr$_6$ at room temperature.

*b) K-Edge Sn XAS spectroscopy.* K-edge X-ray absorption spectra were collected on pressed pellets (6mm diameter, 2mm thickness) of CsSnX$_3$, Cs$_2$SnX$_6$, SnI$_2$ and SnI$_4$. The samples were prepared by mixing 17 mg of the Sn compounds with 85 mg of dry BN following by homogenization of the powders using mortar and pestle. The homogeneous mixtures were pressed in a hydraulic press using a pressure of ~0.35 GPa to form dense pellets. The pellets were measured at Advanced Photon Source (Beamline 20-ID) in the 29050-30427 keV range using a step of 5 eV for the pre-edge region, 1.5 eV for the edge region and 0.05eV for the post-edge region with an integration time of 1s. The spectra were measured against a Sn foil standard. The data were processed and analyzed using Demeter.[5]

*c) $^{119}$Sn Mössbauer Spectroscopy.* $^{119}$Sn Mössbauer spectra were collected on powder samples using a constant-acceleration Mössbauer spectrometer equipped with a Ca$^{119m}$SnO$_3$ source kept at room temperature (RT = 291-293 K) and a variable-temperature cryostat (Thor Cryogenics). The spectrometer was calibrated with metallic α-Fe at RT. Analyses of the spectra were performed using a least-squares Mössbauer fitting program [Douvalis, A. P.; Polymeros, A.; Bakas, T. J. Phys.: Conf. Ser. 2010, 217, 012014.] and the isomer shift (IS) values of the components used to fit the spectra are given relative to SnO$_2$ at RT.

**Table S1.** Selected $^{119}$Sn Mössbauer parameters for CsSnX$_3$, Cs$_2$SnX$_6$, SnI$_2$ and SnI$_4$

| Sample | Isomer Shift (mm/s) | Γ/2 | Quadrupole splitting | Area | FOS |
|---|---|---|---|---|---|

| | | (mm/s) | (mm/s) | (%) | |
|---|---|---|---|---|---|
| SnI$_4$ | 1.48 | 0.11 | 0.00 | 100 | Sn$^{IV}$ in α-SnI$_4$ |
| SnI$_2$ | -0.01 | 0.36 | 0.53 | 83 | Sn$^{IV}$ in α-SnO$_2$ |
| | 3.87 | 0.31 | 0.17 | 17 | Sn$^{II}$ in α-SnI$_2$ |
| Cs$_2$SnI$_6$ | 1.32 | 0.39 | 0.00 | 100 | Sn$^{IV}$ in α-Cs$_2$SnI$_6$ |
| Cs$_2$SnCl$_6$ | 0.46 | 0.39 | 0.12 | 100 | Sn$^{IV}$ in α-Cs$_2$SnCl$_6$ |
| Cs$_2$SnBr$_6$ | 0.83 | 0.37 | 0.17 | 100 | Sn$^{IV}$ in α-Cs$_2$SnBr$_6$ |
| CsSnBr$_3$ | 0.044 | 0.33 | 0.51 | 5 | Sn$^{IV}$ in α-SnO$_2$ |
| | 3.90 | 0.37 | 0.15 | 95 | Sn$^{II}$ α-CsSnBr$_3$ |

## S3 Theoretical Calculations

The *ab initio* calculations were performed by using plane-wave pseudopotential approach within density functional theory (DFT) as implemented in the Viena Ab Initio Simulation Package (VASP).[6,7] We've used the projected augmented wave pseudopotentials[7] with a valence configuration as: Cs($5s^25p^66s^1$), Sn($5s^22p^2$), F($2s^22p^5$), Cl($3s^23p^5$), Br($4s^24p^5$) and I($5s^25p^5$). The generalized gradient approximation formulated by Perdew, Burke, and Ernzerhof (PBE)[8] is used as exchange correlation functional. Experimental lattice parameters and atomic positions were used through the calculations. These are very good models since we are mostly interested on the change density differences. K-point sets of 10x10x10 for 113 compounds and 6x6x6 for 216 compounds were used. To reduce the self-interaction error of DFT for band gaps and band structured calculations, we used the Heyd-Scuseria-Ernzerhof (HSE) hybrid functional approach with standard 25% exact Fock exchange included.[9]

## S4 Structural properties for 113 and 216 compounds

In order to form the 216 compound, Sn atoms are removed from 113 in such a way that a stacking between planes of vacancies and of Sn atoms are formed in the (111) direction, as shown in Figure S1. The removal of 50% of the atoms leads to strong relaxations, significantly decreasing the bond length between Sn and I atoms and thus the volume of the [SnI6] octahedron. Table SI shows measured lattice parameters for the halide compounds. We note that there is a -8% difference in the Sn-I bond length for the Cs$_2$SnI$_6$ compared with CsSnI$_3$. On the other hand, the Shannon-Prewitt effective ionic radius for Sn is 0.69 Å for Sn (IV+) and 0.96 Å for Sn(II+), with a much larger difference (-28%). This raises interesting questions on how the physical charges relate to the FOS of the Sn atoms in these perovskite compounds.

Table S2: experimental structural parameters for halides.

| $A_nMX_{3n}$ | $Cs_2SnI_6$ | $CsSnI_3$ (500K) | $Cs_2SnBr_6$ | $CsSnBr_3$ | $Cs_2SnCl_6$ | $CsSnCl_3$[10] |
|---|---|---|---|---|---|---|
| Structure | cubic | Cubic | cubic | cubic | cubic | cubic |
| space group | *Fm-3m* | *Pm-3m* | *Fm-3m* | *Pm-3m* | *Fm-3m* | *Pm-3m* |
| a (Å) | 11.6276(9) | 6.20589(1) | 10.8265(2) | 5.7956(7) | 10.3721(2) | 5.504 |
| d(x-x) (Å) | 4.0332(6) | 4.38819(3) | 3.713(1) | 4.0981(5) | 3.483(3) | 3.892 |
| d(m-x) (Å) | 2.8519(6) | 3.10295(2) | 2.625(1) | 2.8978(4) | 2.463(3) | 2.752 |
| d(a-x) (Å) | 4.1113(3) | 4.38819(3) | 3.8286(1) | 4.0981(5) | 3.6694(1) | 3.892 |

**S5 Band structures for 113 and 216 compounds calculated with hybrid functionals**

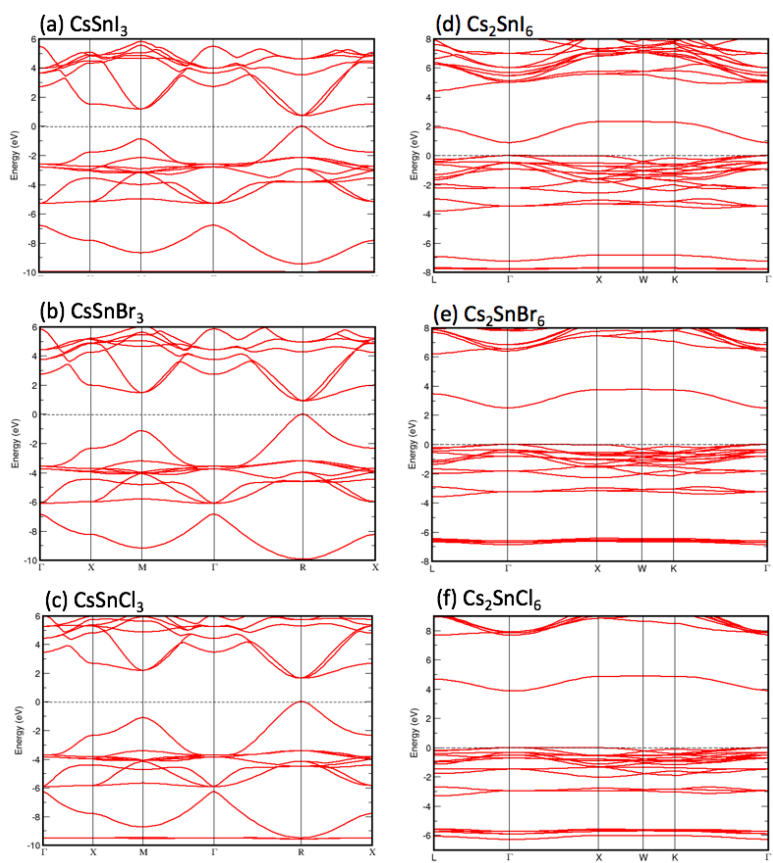

**Figure S6.** Band structures of halide perovskites. Dotted lines indicate the valence band maximum. All calculations were performed using hybrid functionals and the experimental lattice parameters.

**S6 Charge density difference at the 216 lattice parameter.**

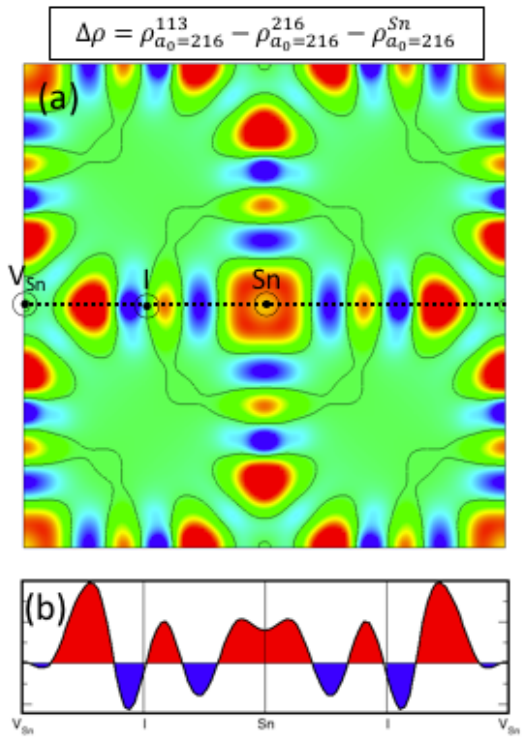

**Fig. S7:** Charge density difference at the 216 compound lattice parameter.

**S7 Band diagram for the fluorides**

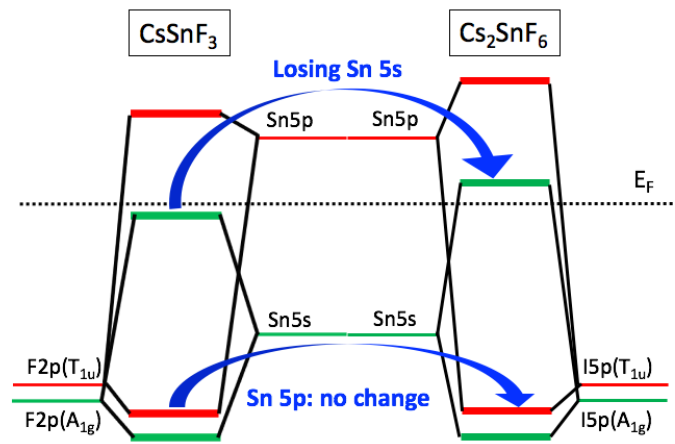

**Fig. S8:** Energy diagram for fluorides showing the inversion of the energetic order of F2p and Sn5s levels, minimizing the SRR effect.

**S8. The fully ionic limit: CsSrI$_3$**

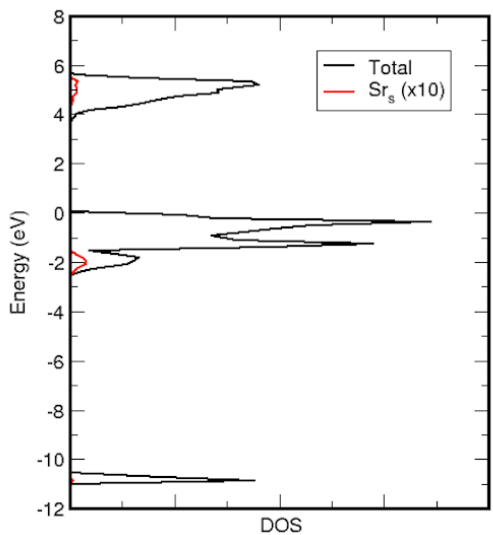

**Figure S9.** Projected density of states for CsSrI$_3$. The Sr s contribution is very small, indicating th fully ionic, FOS IV limit.

**S9. The relevance of the SnX6 molecule to the discussion above**

The progression from 113 to 216 and then to the isolated $[SnI_6]^{2-}$ molecule reveals the trend of the bands becoming narrower and more localized as the SnI6 units are more separated: By comparing the total DOS of the two iodides, we find that CsSnI3 has broader DOS bands, while $Cs_2SnI_6$ has sharper DOS peaks (see Fig 2a and b). This is because the main contribution to states near the Fermi level is from the $SnI_6$ octahedron, which has higher packing density in the 113 compounds (with corner sharing octahedra) whereas in the 216 compounds the octahedra are discrete and isolated. We also considered the electronic structure of the hypothetical molecular limit of a $[SnI6]^{2-}$ octahedron as comparison. The results (Fig. S8) illustrate that the dominant hybridization in 216 is inside each $[SnI_6]^{2-}$ octahedron**.**

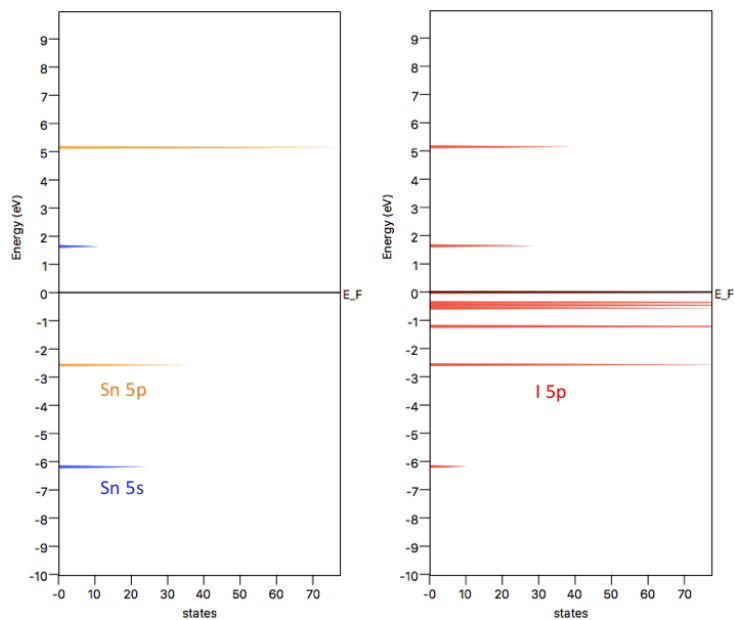

Fig. S10: projected DOS of [SnI6]2- molecule.